\documentclass[11pt]{article}

\usepackage{amsmath,amssymb,amsthm,amscd,latexsym}
\usepackage[toc,title,titletoc]{appendix}
\usepackage{cite}
\usepackage{graphicx,epsfig}
\usepackage[breaklinks,
colorlinks=true,
linkcolor=red,
urlcolor=blue]{hyperref}
\usepackage{breakurl}

\RequirePackage{color}

\allowdisplaybreaks[1]

\def\abs#1{\left| #1 \right|}

\def\nk{n_{\rm b}}

\def\rfr#1{eq. (\ref{#1})}

\def\derp#1#2{\rp{\partial{#1}}{\partial{#2}}}

\def\virg#1{``#1''}

\def\eqi{\begin{equation}}
\def\eqf{\end{equation}}
\def\eqia{\begin{eqnarray}}
\def\eqfa{\end{eqnarray}}
\def\Om{\mathit{\Omega}}
\def\rp#1#2{{#1\over#2}}
\def\lb#1{\label{#1}}

\def\ton#1{\left(#1\right)}
\def\qua#1{\left[#1\right]}

\def\ang#1{\left\langle #1\right\rangle}

\begin{document}

\title{On a recent preliminary study for the measurement of the Lense-Thirring effect with the Galileo satellites}

\author{L. Iorio\\ Ministero dell'Istruzione, dell'Universit$\grave{\textrm{a}}$ e della Ricerca (M.I.U.R.)-Istruzione \\ Fellow of the Royal Astronomical Society (F.R.A.S.)\\ Viale Unit$\grave{\textrm{a}}$ di Italia 68, 70125, Bari (BA), Italy}

\maketitle

\begin{abstract}
 It has recently been  proposed to combine the node drifts of the future constellation of 27 Galileo spacecraft together with those of the existing LAGEOS-type satellites  to improve the accuracy of the past and ongoing tests of the Lense-Thirring (LT) effect  by removing the bias of a larger number of even zonal harmonics $J_{\ell}$ than either done or planned so far. Actually, it seems a difficult goal to be achieved realistically  for a number of reasons. First, the LT range signature of a Galileo-type satellite is as small as $0.5$ mm over 3-days arcs, corresponding to a node rate of just $\dot\Om_{\rm LT}=2$ milliarcseconds per year (mas yr$^{-1}$). Some tesseral  and sectorial ocean tides such as $K_1,K_2$ induce long-period harmonic node perturbations with  frequencies which are integer multiples  of the extremely slow Galileo's node rate $\dot\Om$ completing a full cycle in about 40 yr. Thus, over time spans $T$ of some years they would act as superimposed semi-secular aliasing trends. Since the coefficients of the $J_{\ell}$-free multisatellite linear combinations are determined only by the  semimajor axis $a$, the eccentricity $e$ and the inclination $I$, which are nominally equal for all the Galileo satellites, it is not possible to include all of them. Even by using only one Galileo spacecraft together with the LAGEOS family would be unfeasible because the resulting Galileo coefficient would be $\gtrsim 1$, thus enhancing  the aliasing impact of the uncancelled non-conservative and tidal perturbations.
\end{abstract}

\leftline{Keywords:}
\leftline{Experimental studies of gravity;}
\leftline{Experimental tests of gravitational theories;}
\leftline{Satellite orbits;}
\leftline{Earth tides;}
\leftline{Harmonics of the gravity potential field}
\bigskip
\leftline{PACS: 04.80.-y; 04.80.Cc; 91.10.Sp; 91.10.Tq; 91.10.Qm}


\section{Introduction}\lb{Introduzione}
According to the General Theory of Relativity (GTR), the orbital plane of a test particle in geodesic motion around a rotating body of mass $M$, equatorial radius $R$ and proper angular momentum $S$ is secularly dragged in the same direction as that of the body's rotation according to
\eqi\dot\Om_{\rm LT} = \rp{2GS}{c^2 a^3\ton{1-e^2}^{3/2}}.\lb{LTnode}\eqf
 In \rfr{LTnode}, $\Om$ is the longitude of the ascending node of the satellite's orbital plane, $G$ is the Newtonian gravitational constant, $c$ is the speed of light in vacuum, and $a,e$ are the satellite's semimajor axis and eccentricity, respectively. The node precession of \rfr{LTnode} is one of the manifestations\footnote{Also the argument of pericenter $\omega$ and the longitude of pericenter $\varpi$ of a test particle undergo  secular relativistic  precessions  due to $S$ \cite{LT18}. They will not be treated here.}  of the so-called Lense-Thirring  effect \cite{LT18}, taking place in the gravitomagnetic field \cite{libro2, Rin01, libro} of a massive spinning object. In general, it is due to the off-diagonal components $g_{0i},\ i=1,2,3$ of the spacetime metric generated by a stationary mass-energy distribution. Gravitomagnetism affects in various ways the motion of test particles, the precession of gyroscopes and the propagation of electromagnetic waves  \cite{2002NCimB.117..743R}.
Recent years have seen increasing efforts to measure the Lense-Thirring orbital drag with natural and artificial probes in the Solar System; for a general review, see \cite{2011Ap&SS.331..351I} and references therein. As far as the Earth is concerned, a history of the attempts made so far with man-made satellites can be found in \cite{2013CEJPh..11..531R}. The gravitomagnetic gyroscope precession, known also as the Pugh-Schiff effect \cite{Pugh59, Schiff60}, was recently measured in the gravitational field of the spinning Earth with the dedicated space-based Gravity Probe B (GP-B) mission at a claimed $19\%$ accuracy \cite{2011PhRvL.106v1101E, 2013NuPhS.243..172W}.

In order to improve the accuracy of the performed and ongoing attempts to measure the Lense-Thirring effect with  the Satellite Laser Ranging (SLR) technique \cite{2002AdSpR..30..135P, 2004hlta.book.2531W} applied to the geodetic LAGEOS-like satellites \cite{2011EPJP..126...72C}, the authors of \cite{2013arXiv1311.3917M}  recently  proposed  to use the future constellation of 27 navigation spacecraft of the planned European Global Navigation Satellite System Galileo by suitably combining their data with those of LAGEOS, LAGEOS II and LARES. Two test satellites, named GIOVE$-$A and GIOVE$-$B, were launched as part of the Galileo in Orbit Validation Element (GIOVE) \cite{2006ESABu.127...62B, gioveb}.

In this paper, we  critically discuss such a proposal. In Section \ref{grossa}, we look at some potential issues of it. The actual possibility of extracting the Lense-Thirring signature from the station-satellite SLR range measurements of a typical Galileo satellite is the subject of Section \ref{rangio}. The impact of the mismodeling in the static component of the first even zonal harmonic of the geopotential  on the range and on the node of a Galileo-type spacecraft is treated in Section \ref{geopo}. Section \ref{maree} is devoted to the bias of certain semi-secular tidal orbital perturbations on the node of a navigation spacecraft of the type considered here. The consequences of including the nodes of the members of the Galileo constellation in linear combinations with the nodes of the SLR targets of the LAGEOS family is dealt with in Section \ref{combiz}. Finally, in Section \ref{fine} we offer our conclusions.
\section{Potential issues implied by the use of Galileo}\lb{grossa}
The relevant orbital parameters of LAGEOS, LAGEOS II, LARES and of a typical spacecraft of the Galileo family are listed in Table \ref{tavola1}.
\begin{table*}[ht!]
\caption{Semimajor axis $a$, eccentricity $e$, inclination $I$ to the Earth's equator, period $P_{\Om}$ of the node, and Lense-Thirring node precessions $\dot\Om_{\rm LT}$, in milliarcseconds per year (mas yr$^{-1}$), of the three satellites of the LAGEOS family (LAGEOS, LAGEOS II, LARES), and of a member of the planned Galileo constellation.
}\label{tavola1}
\centering
\bigskip
\begin{tabular}{llllll}
\hline\noalign{\smallskip}
  Satellite & $a$ (km)  & $e$  & $I$ (deg) & $P_{\Om}$ (yr) & $\dot\Om_{\rm LT}$ (mas yr$^{-1}$)  \\
\noalign{\smallskip}\hline\noalign{\smallskip}
LAGEOS & $12270$ & $0.0045$  & $109.9$ & $2.87$ & $30.7$ \\
LAGEOS II & $12163$ & $0.014$ & $52.65$ & $1.56$  & $31.5$ \\
LARES & $7828$ & $0.0007$ & $69.5$ & $0.57$  & $118.4$ \\
Galileo & $29600$ & $0.0$ & $56$ & $38.1$  & $2.2$ \\
\noalign{\smallskip}\hline\noalign{\smallskip}
\end{tabular}
\end{table*}
The period of the node is mainly determined by the first  even ($\ell=2$) zonal ($m=0$) harmonic coefficient $J_2 = -\sqrt{5}\ {\overline{C}}_{2,0}$ of the multipolar expansion of the Earth's gravitational potential \cite{Heis67} causing the secular precession
\eqi\dot\Om_{J_2} = -\rp{3}{2}\nk\ton{\rp{R}{a}}^2\rp{J_2\cos I}{\ton{1-e^2}^2},\lb{J2node}\eqf
where  $\nk=\sqrt{GM a^{-3}}$ is the satellite's Keplerian mean motion, and ${\overline{C}_{2,0}}$ is the normalized Stokes coefficient of degree $\ell=2$ and order $m=0$. As we will see in Section \ref{maree}, the duration $P_{\Om}$ of the satellite's node rotation with respect to the time interval $T$ of the data analysis is important since there are certain long-period harmonic orbital perturbations with the same frequency of the node itself. If $T \ll P_{\Om}$ and their mismodeled amplitudes are non-negligible, such signals may represent a source of major systematic bias.
Another remarkable feature emerging from Table \ref{tavola1} is the smallness of the Lense-Thirring node drag for a Galileo-type spacecraft.
\subsection{The actual detectability of the Lense-Thirring effect in the Galileo station-satellite range}\lb{rangio}
The direct observable with the SLR technique, applied also to the Galileo spacecraft each of which will carry a laser retroreflector array\footnote{See \url{http://ilrs.gsfc.nasa.gov/missions/satellite_missions/current_missions/ga01_reflector.html} on the WEB.}, is the  
two-way time-of-flight measurement \cite{librus}. This is then corrected for the additional delay due to the atmosphere, satellite centre-of-mass, the Shapiro delay, etc.
As a consequence, a station-satellite range $\rho$ is obtained. Thus, it is important to look at the impact of the Earth's gravitomagnetic field on such a quantity for Galileo-type satellites. In Figure \ref{LTrange} we show the simulated Lense-Thirring range shift $\Delta\rho$ over a 3-days arc, as adopted in \cite{2013arXiv1311.3917M},  for, say, the Graz SLR station and a given choice of initial conditions for the node and for the mean anomaly of a typical Galileo spacecraft.
\begin{figure*}
\centering
\begin{tabular}{c}
\epsfig{file=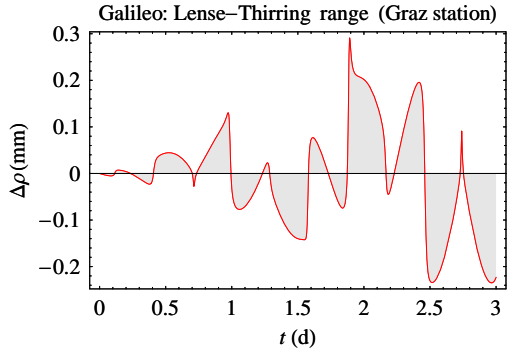,width=0.60\linewidth,clip=}\\
\end{tabular}
\caption{Simulated Lense-Thirring station-satellite range signal $\Delta\rho$ for a Galileo spacecraft and the SLR Graz station over a 3-days arc. By varying the initial conditions $\Om_0,{\mathcal{M}}_0$ for the node  and  the mean anomaly of Galileo does not induce appreciable changes in the overall pattern, especially as far as the magnitude of the signal is concerned.}\lb{LTrange}
\end{figure*}
It turns out that the peak-to-peak amplitude is as small as $0.5$ mm.
The present-day  accuracy in reconstructing the  orbits of the test satellites GIOVE-A and GIOVE-B with SLR data, is at the $\approx 1-15$ cm level, in the RMS sense, for arcs spanning a few days \cite{gnss, 2011JGeod..85..357S}.
\subsection{The impact of the mismodeling in the static part of the geopotential}\lb{geopo}
A major source of systematic uncertainty in measuring the Lense-Thirring effect is represented by the even ($\ell=2,4,6,\ldots$) zonal ($m=0$) harmonics of the static part of the geopotential \cite{Heis67} which affect the orbit of a satellite with signatures which are qualitatively equal to the gravitomagnetic one, but quantitatively much larger. The largest precession is due to $J_2$; it is explicitly displayed in \rfr{J2node}.

As far as Galileo is concerned, in Figure \ref{rangeJ2} we depict the simulated residual range signal induced by a mismodeling of $\delta J_2$ as large as  the formal, statistical error  \eqi\sigma_{{\overline{C}}_{2,0}}=1.2\times 10^{-10}\lb{sigma}\eqf
released in the recent GOCE-only JYY$\_$GOCE02S model \cite{2012AdSpR..50..371Y, 2013StGG...57..174Y}.
Interestingly, \rfr{sigma} is of the same order of magnitude of the realistic one computed in \cite{2012JHEP...05..073I} by comparing different global gravity field solutions on the basis of the independent approach in \cite{2012JGeod..86...99W}.
\begin{figure*}
\centering
\begin{tabular}{c}
\epsfig{file=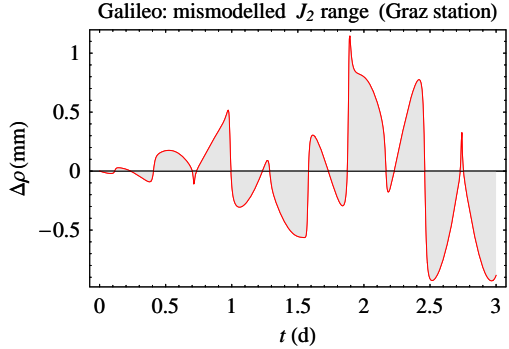,width=0.60\linewidth,clip=}\\
\end{tabular}
\caption{Simulated  station-satellite residual range signal $\Delta\rho$ due to the mismodeling in $J_2$ for a Galileo spacecraft and the SLR Graz station over a 3-days arc. By varying the initial conditions $\Om_0,{\mathcal{M}}_0$ for the node and  the mean anomaly of Galileo does not induce appreciable changes in the overall pattern, especially as far as the magnitude of the signal is concerned. The (formal) error $\sigma_{{\overline{C}}_{2,0}}=1.2\times 10^{-10}$ from the GOCE-only JYY$\_$GOCE02S model \cite{2012AdSpR..50..371Y, 2013StGG...57..174Y} was adopted as representative of the uncertainty $\delta J_2$.}\lb{rangeJ2}
\end{figure*}
Its peak-to-peak amplitude is as large as 2 mm, which is 4 times larger than the expected Lense-Thirring signal in Figure \ref{LTrange}.
From the point of view of a sensitivity analysis, useful to asses the systematic part of a realistic error budget, it is useful to compare the mismodeled precession of \rfr{J2node} for different evaluations of $\delta J_2$ to the Lense-Thirring nominal rate of \rfr{LTnode}. The result  is summarized in Table \ref{tavola2}, where different recent global Earth's gravity field models are used. It must be stressed that all the released errors for ${\overline{C}}_{2,0}$ are formal; the realistic uncertainties can be much larger, as shown  in \cite{2012JHEP...05..073I, 2012JGeod..86...99W}.
\begin{table*}[ht!]
\caption{Percent systematic bias due to the mismodeling in the static component of the first even zonal harmonic on the Lense-Thirring node precession of a typical Galileo spacecraft. For calculating $\partial\dot\Om_{J_2}/\partial J_2$, \rfr{J2node} was adopted. The acronym IWM stands for Iorio-Wagner-McAdoo \cite{2012JHEP...05..073I, 2012JGeod..86...99W}.
}\label{tavola2}
\centering
\bigskip
\begin{tabular}{lll}
\hline\noalign{\smallskip}
  $\ton{\derp{\dot\Om_{J_2}}{J_2}}\rp{\delta J_2}{\dot\Om_{\rm LT}}\ (\%)$ &  $\delta{\overline{C}}_{2,0}$ & Model \\
\noalign{\smallskip}\hline\noalign{\smallskip}
350 & $1.0\times 10^{-10}$ & IWM (calibrated) \cite{2012JHEP...05..073I, 2012JGeod..86...99W}\\
396 & $1.2\times 10^{-10}$ & JYY$\_$GOCE02S ($\sigma_{{\overline{C}}_{2,0}}$, formal) \cite{2012AdSpR..50..371Y, 2013StGG...57..174Y}\\
2 & $6\times 10^{-13}$ & GOGRA02S ($\sigma_{{\overline{C}}_{2,0}}$, formal) \cite{2012AdSpR..50..371Y, 2013StGG...57..174Y} \\
29 & $9\times 10^{-12}$ & ITG-Goce2 ($\sigma_{{\overline{C}}_{2,0}}$, formal) \cite{ITGgoce2} \\
\noalign{\smallskip}\hline\noalign{\smallskip}
\end{tabular}
\end{table*}
From Table \ref{tavola2} it turns out that, even if it was possible to detect the Lense-Thirring effect in the station-satellite ranges, the lingering uncertainty in the first even zonal makes the prospect of using only one Galileo node unfeasible. This serious drawback may be, in principle, cured by combining a Galileo node with the nodes of other SLR targets such as the satellites of the LAGEOS family \cite{2013arXiv1311.3917M}.
\subsection{The solid and ocean tides}\lb{maree}
Solid (or body) and ocean tides are able to induce long-term orbital perturbations on the motion of a satellite moving in the free space potential of a tidally distorted Earth \cite{2001CeMDA..79..201I, 2002CeMDA..82..301K}.

Table \ref{tavola1} shows that the node of a typical Galileo spacecraft is very slow, completing a full cycle in about 40 yr. This fact introduces a further source of major systematic bias. Indeed, some tidal lines, like $K_1, K_2$ in  Darwin's nomenclature \cite{2010spsg.book.....D}, affect the motion of a satellite with long-period harmonic orbital perturbations having frequencies equal to integer multiples of the satellite's node frequency $\dot\Om$ itself \cite{2001CeMDA..79..201I}. Over observational time spans $T$ necessarily limited to some years, they can resemble almost linear trends superimposed to the relativistic one of interest, thus negatively impacting its possible recovery. Such a risk is more pronounced for the ocean tides which, if on the one hand have nominal amplitudes smaller than the solid ones, on the other hand  are known less accurately.

The node shifts $\Delta\Om$ due to the solid and ocean ($\ell=2,p=1,q=0$, prograde\footnote{They are the Westwards waves. The Eastwards retrograde waves due to the non-equilibrium pattern of the ocean tidal bulge, denoted with a  superscript \virg{-}, do not cause long-period orbital perturbations.}) components of the tesseral ($m=1$) $K_1$ tide, with Doodson number (165.555), are \cite{2001CeMDA..79..201I}
\begin{align}
\Delta\Om^{(\rm solid)}_{K_1} \lb{K1sol}&= {\mathcal{A}}_{K_1}^{(\rm solid)}\sin\ton{\dot\Om t + \Om_0-\delta_{2,1,K_1}}, \\ \nonumber \\
\Delta\Om^{(\rm ocean)}_{K_1} \lb{K1oc}&= -{\mathcal{A}}_{K_1}^{(\rm ocean)}\cos\ton{\dot\Om t + \Om_0-\varepsilon^{+}_{2,1,K_1}}.
\end{align}
In \rfr{K1sol}-\rfr{K1oc}, $\dot\Om$ is the total node frequency, mainly determined by $J_2$ according to \rfr{J2node}, $\Om_0$ is the initial value of the satellite's node, the dimensionless, frequency\footnote{It is referred to the tidal line.}-dependent parameters $\delta_{2,1,K_1},\varepsilon^{+}_{2,1,K_1}$ are the solid and ocean phase lag angles\footnote{$\delta_{2,1,K_1}$ is due to the anelasticity of the  mantle, while $\varepsilon^{+}_{2,1,K_1}$ is caused by the complex hydrodynamics of the oceans (inertia of the running fluid elements, dissipative phenomena and non-linear interactions among tidal and other ocean currents) \cite{IERS010}.}, respectively \cite{2001CeMDA..79..201I}. The dimensionless tidal amplitudes $\mathcal{A}$ are
\begin{align}
{\mathcal{A}}_{K_1}^{(\rm solid)} \lb{AK1sol}& = -\sqrt{\rp{15}{2\pi G M a^7}}\rp{ g R^3 k^{(0)}_{2,1,K_1} H_2^1(K_1) \cos 2I\csc I}{4\ton{1-e^2}^2\dot\Om }, \\ \nonumber \\
{\mathcal{A}}_{K_1}^{(\rm ocean)} \lb{AK1oc}& = -\sqrt{\rp{G}{M a^7}}\rp{ 6\pi \rho_{\rm w} R^4 C_{2,1,K_1}^{+}\ton{1+k^{'}_2}\cos 2 I\csc I}{5\ton{1-e^2}^2\dot\Om }.
\end{align}
 As far as the meaning of the geophysical and hydrodynamical symbols entering \rfr{AK1sol}-\rfr{AK1oc} is concerned, see \cite{2001CeMDA..79..201I, IERS010} and references therein. Here, we briefly recall that $g$ is the terrestrial acceleration of gravity at the equator (m s$^{-2}$), $k^{(0)}_{2,1,K_1}$ is the dimensionless frequency-dependent Love number\footnote{The contributions of ellipticity and of the Coriolis force, accounted for by $k^{+}_{2,1} = -0.00080$, are neglected because of its smallness \cite{IERS010}.} for the $K_1$ tidal constituent, $\rho_{\rm w}$ is the volumetric ocean water density (kg m$^{-3}$), $k_2^{'}$ is the dimensionless load Love number, $H_2^1(K_1), C^{+}_{2,1,K_1}$ are the frequency-dependent solid and ocean tidal heights (m), respectively.
\begin{table*}[ht!]
\caption{Relevant geophysical and hydrodynamical Earth's parameters for the $f=K_1$ and $f=K_2$ tides. The figures for the ocean tidal heights $C_{\ell,m,f}^{+}$ and phase lags $\varepsilon_{\ell,m,f}^{+}$ along with their errors $\sigma_{C_{\ell,m,f}^{+}}, \sigma_{\varepsilon_{\ell,m,f}^{+}}$  are retrieved from the EGM96 model \cite{EGM96}.
}\label{tavola3}
\centering
\bigskip
\begin{tabular}{lll}
\hline\noalign{\smallskip}
  Parameter & Numerical value  & Units  \\
\noalign{\smallskip}\hline\noalign{\smallskip}
$\rho_{\rm w}$ & $1025$ & kg.m$^{-3}$ \\
$g$ & $9.7803278$ & m.s$^{-2}$\\
$k^{(0)}_{2,1,K_1}$ & $0.257$ & - \\
$H_2^1$ ($K_1$) & $0.3687012$ & m \\
$\delta_{2,1,K_1}$ & $-18.36$ & deg \\
$k_2^{'}$ & $-0.3075$ & - \\
$C_{2,1,K_1}^{+}$ & $0.0283\pm 0.0012$ & m \\
$\varepsilon_{2,1,K_1}^{+}$ & $320.6 \pm 2.2$ & deg \\
$k^{(0)}_{2,2}$ & $0.301$ & - \\
$H_2^2$ ($K_2$) & $0.0799155$ & m \\
$\delta_{2,2}$ & $-14.15$ & deg \\
$C_{2,2,K_2}^{+}$ & $0.0027\pm 0.0003$ & m \\
$\varepsilon_{2,2,K_2}^{+}$ & $328.4 \pm 5.7$ & deg \\
\noalign{\smallskip}\hline\noalign{\smallskip}
\end{tabular}
\end{table*}
As a consequence of \rfr{K1sol}-\rfr{K1oc}, with \rfr{AK1sol}-\rfr{AK1oc}, the ratios of the mean tidal and gravitomagnetic node shifts, averaged over a time interval $T$, are
\begin{align}
\rp{\ang{\Delta\Om^{(\rm solid)}_{K_1}}}{\ang{\Delta\Om_{\rm LT}}} & = \rp{4{\mathcal{A}}_{K_1}^{(\rm solid)} \qua{\sin\ton{\rp{\dot \Om}{2} T}\sin\ton{\rp{\dot\Om}{2} T+\Om_0-\delta_{2,2,K_1}}}  }{\dot\Om\dot \Om_{\rm LT}T^2}, \\ \nonumber \\
\rp{\ang{\Delta\Om^{(\rm ocean)}_{K_1}}}{\ang{\Delta\Om_{\rm LT}}} & = -\rp{2{\mathcal{A}}_{K_1}^{(\rm ocean)}\qua{\sin\ton{\varepsilon_{2,1,K_1}^{+} -\Om_0} +\sin\ton{\dot\Om t + \Om_0 -\varepsilon_{2,1,K_1}^{+}} }  }{\dot\Om\dot \Om_{\rm LT}T^2},
\end{align}

The node shifts $\Delta\Om$ due to the solid and ocean ($\ell=2,p=1,q=0$, prograde) components of the sectorial ($m=2$) $K_2$ tide, with Doodson number (275.555), are \cite{2001CeMDA..79..201I}
\begin{align}
\Delta\Om^{(\rm solid)}_{K_2} &= {\mathcal{A}}_{K_2}^{(\rm solid)}\sin\qua{2\ton{\dot\Om t + \Om_0}-\delta_{2,2}}, \\ \nonumber \\
\Delta\Om^{(\rm ocean)}_{K_2} &= {\mathcal{A}}_{K_2}^{(\rm ocean)}\sin\qua{2\ton{\dot\Om t + \Om_0}-\varepsilon^{+}_{2,2,K_2}},
\end{align}
where\footnote{The Love number $k_{2,m}^{(0)}$ and the lag angle $\delta_{2,m}$ for the sectorial band ($m=2$) do not depend on the tidal constituents, apart from $N_2$ (245.655) and $M_2$ (255.555) \cite{IERS010}.}
\begin{align}
{\mathcal{A}}_{K_2}^{(\rm solid)} & = \sqrt{\rp{15}{2\pi GM a^7}}\rp{ g R^3 k^{(0)}_{2,2} H_2^2(K_2) \cos I}{8\ton{1-e^2}^2\dot\Om }, \\ \nonumber \\
{\mathcal{A}}_{K_2}^{(\rm ocean)} & = \sqrt{\rp{G}{M a^7}}\rp{ 6\pi\rho_{\rm w} R^4 C_{2,2,K_2}^{+}\ton{1+k^{'}_{2}}\cos I}{5\ton{1-e^2}^2\dot\Om }.
\end{align}
Thus, the ratios of the mean tidal and gravitomagnetic node shifts, averaged over a time interval $T$, are
\begin{align}
\rp{\ang{\Delta\Om^{(\rm solid)}_{K_2}}}{\ang{\Delta\Om_{\rm LT}}} & = \rp{2{\mathcal{A}}_{K_2}^{(\rm solid)} \qua{\sin\ton{\dot \Om T}\sin\ton{\dot\Om T+2\Om_0-\delta_{2,2}}}  }{\dot\Om\dot \Om_{\rm LT}T^2}, \\ \nonumber \\
\rp{\ang{\Delta\Om^{(\rm ocean)}_{K_2}}}{\ang{\Delta\Om_{\rm LT}}} & = \rp{2{\mathcal{A}}_{K_2}^{(\rm ocean)} \qua{\sin\ton{\dot \Om T}\sin\ton{\dot\Om T+2\Om_0-\varepsilon_{2,2,K_2}^{+}}}  }{\dot\Om\dot \Om_{\rm LT}T^2},
\end{align}
As a general remark, the solid tides are better known than the ocean ones, while the  nominal amplitudes of the orbital perturbations of the ocean tides are smaller than those due to the body tides.

Figure \ref{solidtidalplots}, obtained by using Table \ref{tavola3}, displays the percent bias on the Lense-Thirring node rate of a Galileo satellite due to the mismodeling in the Love numbers and phase lag angles of the solid $f=K_1$ and $f=K_2$ constituents. A $0.5\%$ uncertainty in them was assumed \cite{2001CeMDA..79..201I}.
\begin{figure*}
\centering
\begin{tabular}{cc}
\epsfig{file=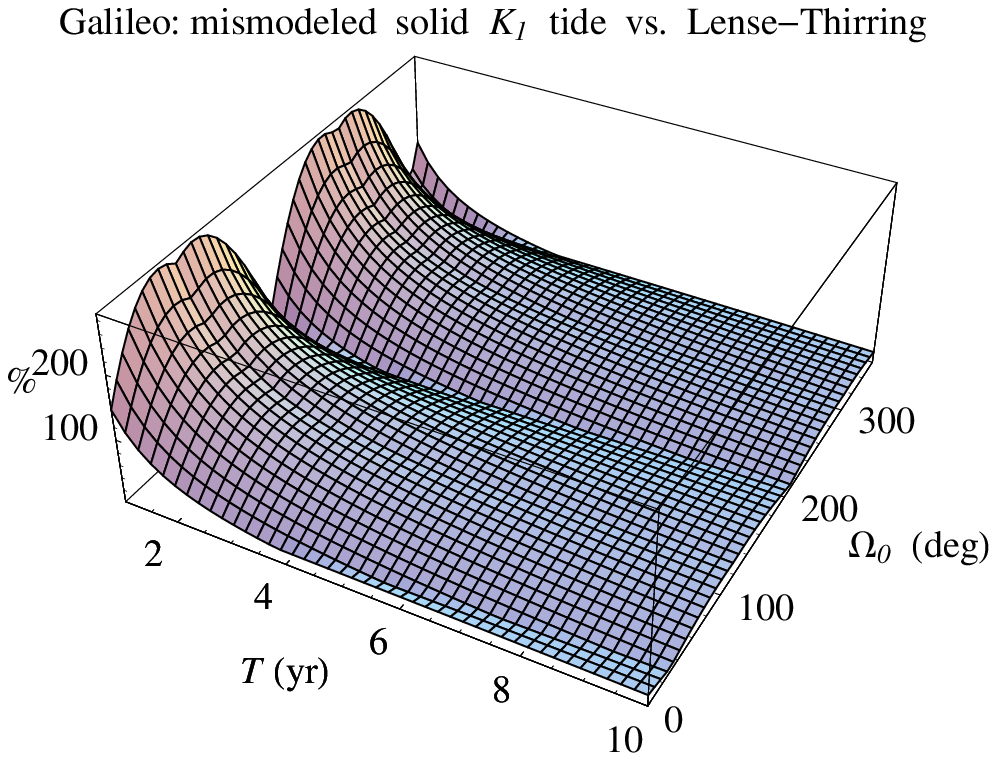,width=0.50\linewidth,clip=} & \epsfig{file=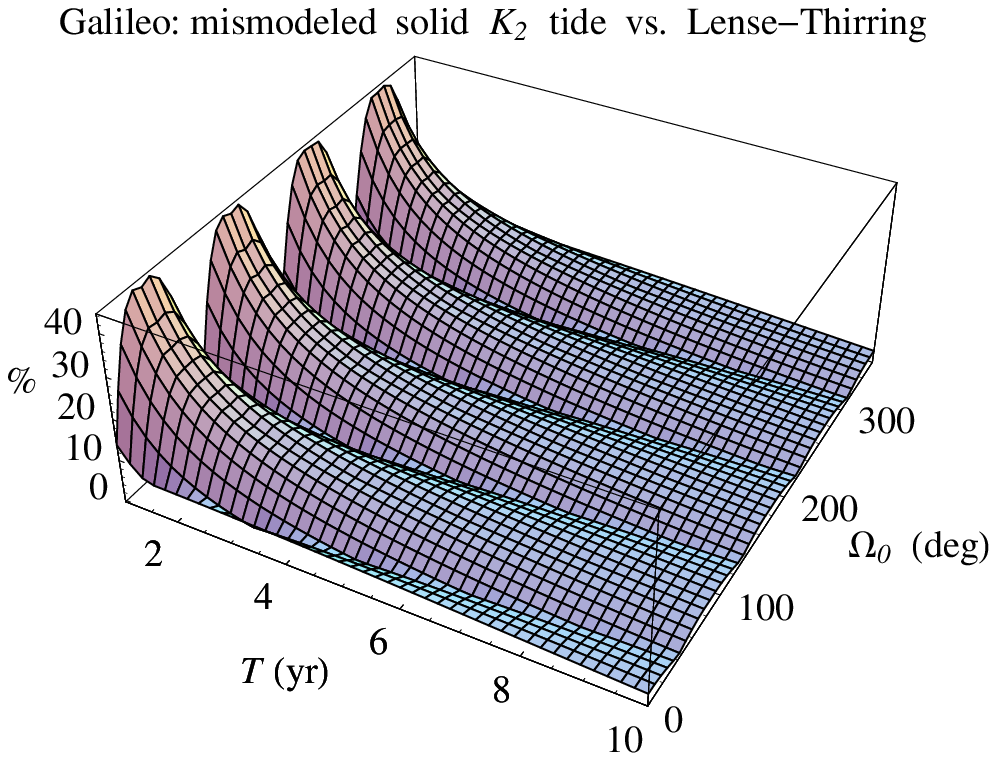,width=0.50\linewidth,clip=}\\
\end{tabular}
\caption{Percent systematic mean error in the Galileo Lense-Thirring node shift induced by the mismodeling in the Love numbers $k^{(0)}_{2,m,f}$ and the phase lag angles $\delta_{2,m,f}$ of the solid $f=K_1$ (left panel) and $f=K_2$ (right panel) tidal constituents as a function of the data analysis time span $T$ and of the initial value $\Om_0$ of the satellite's node. A $0.5\%$  uncertainty level was assumed in $k^{(0)}_{2,m,f}$ and $\delta_{2,m,f}$ \cite{2001CeMDA..79..201I}.}\lb{solidtidalplots}
\end{figure*}
The maximum and minimum percent bias on the Lense-Thirring node precession for a Galileo spacecraft are listed in Table \ref{tavola4} along with the corresponding values for $T,\Om_0$.
\begin{table*}[ht!]
\caption{
Minimum and maximum percent solid tidal bias on the Lense-Thirring node precession for a Galileo satellite (see Figure \ref{solidtidalplots}). Also the corresponding values for the time span $T$ and the initial value $\Om_0$ of the satellite's node are listed. A $0.5\%$ uncertainty was assumed for the Love numbers and the phase lag angles \cite{2001CeMDA..79..201I}.
}\label{tavola4}
\centering
\bigskip
\begin{tabular}{llll}
\hline\noalign{\smallskip}
Tide $f$ & Min/max percent bias $(\%)$ & $\Om_0$ (deg) & $T$ (yr)  \\
\noalign{\smallskip}\hline\noalign{\smallskip}
$K_1$ (min) & $8$ & $29$ & 10 \\
$K_1$ (max) & $282$ & $238$ & 1 \\
$K_2$ (min) & $0.6$ & $40$ & 10 \\
$K_2$ (max) & $43$ & $215$ & 1 \\
\noalign{\smallskip}\hline\noalign{\smallskip}
\end{tabular}
\end{table*}

In Figure \ref{oceantidalplots}, produced by using Table \ref{tavola3}, we plot the percent bias on the Lense-Thirring node rate of a Galileo satellite due to the mismodeling in the tidal height coefficients $C_{\ell,m,f}^{+}$ and the phase lag angles $\varepsilon_{\ell,m,f}^{+}$ of the ocean $f=K_1$ and $f=K_2$ constituents with $\ell=2,p=1,q=0$.
\begin{figure*}
\centering
\begin{tabular}{cc}
\epsfig{file=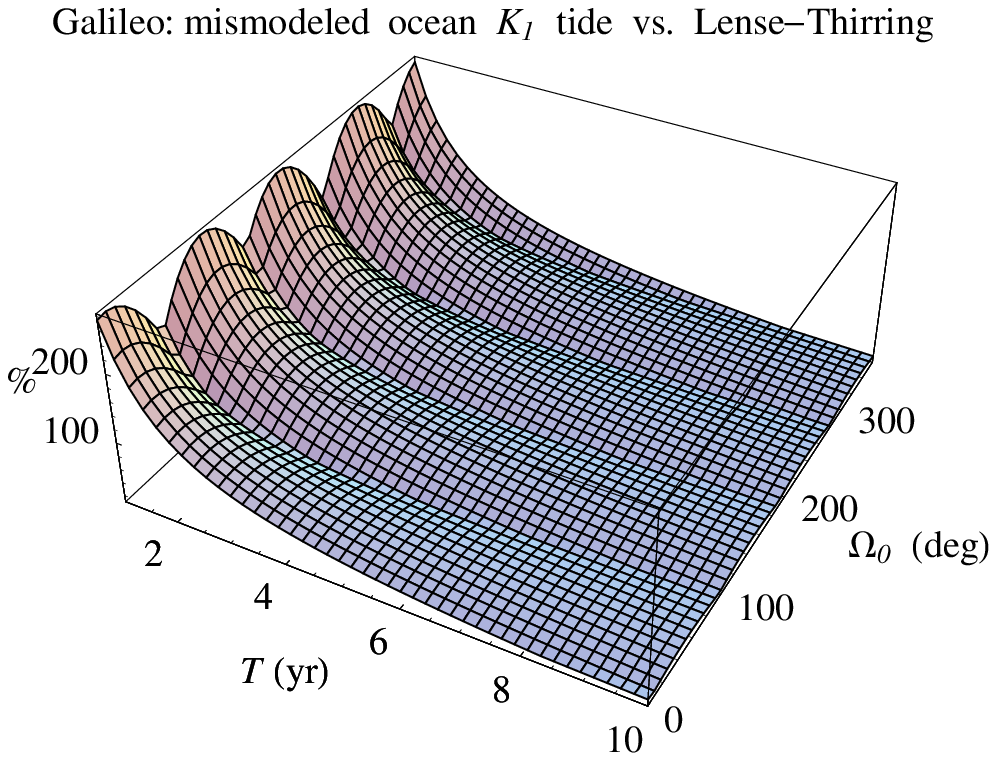,width=0.50\linewidth,clip=} & \epsfig{file=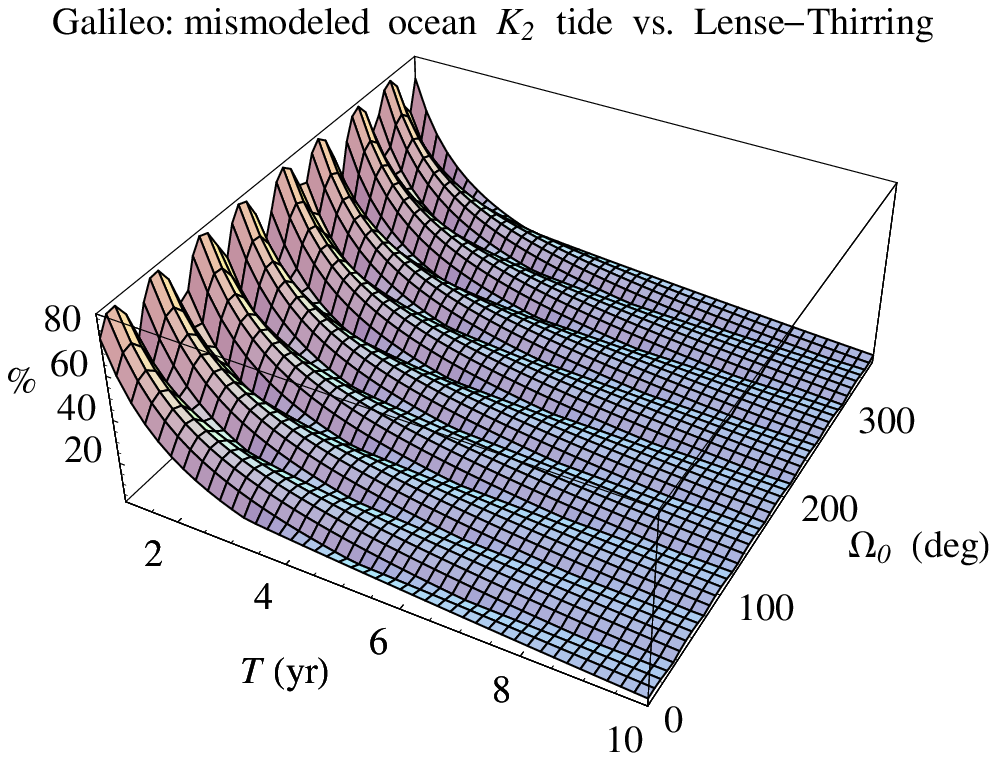,width=0.50\linewidth,clip=}\\
\end{tabular}
\caption{Percent systematic mean error in the Galileo Lense-Thirring node shift induced by the mismodeling in the height coefficients $C_{\ell,m,f}^{+}$ and the phase lag angles $\varepsilon_{\ell,m,f}^{+}$ of the ocean $f=K_1$ (left panel) and $f=K_2$ (right panel) tidal constituents with $\ell=2,p=1,q=0$ as a function of the data analysis time span $T$ and of the initial value $\Om_0$ of the satellite's node. The uncertainties in $C_{\ell,m,f}^{+}$ and $\varepsilon_{\ell,m,f}^{+}$  were taken from the errors in the EGM96 model \cite{EGM96}.}\lb{oceantidalplots}
\end{figure*}
For the uncertainties in such key tidal parameters, we assumed the errors released in the EGM96 model \cite{EGM96}. In Table \ref{tavola5} we show the minimum and the maximum percent bias and the corresponding values for $T,\Om_0$
\begin{table*}[ht!]
\caption{Minimum and maximum percent ocean tidal bias on the Lense-Thirring node precession for a Galileo satellite (see Figure \ref{oceantidalplots}). Also the corresponding values for the time span $T$ and the initial value $\Om_0$ of the satellite's node are listed. The errors  $\sigma_{C_{\ell,m,f}^{+}},  \sigma_{\varepsilon_{\ell,m,f}^{+}}$  by EGM96 \cite{EGM96} were used.
}\label{tavola5}
\centering
\bigskip
\begin{tabular}{llll}
\hline\noalign{\smallskip}
Tide $f$ & Min/max percent bias $(\%)$ & $\Om_0$ (deg) & $T$ (yr)  \\
\noalign{\smallskip}\hline\noalign{\smallskip}
$K_1$ (min) & 16 & $277.8$ & 10 \\
$K_1$ (max) & 277 & $283.1$ & 1 \\
$K_2$ (min) & $3.4$ & $211.4$ & 10 \\
$K_2$ (max) & $85$ & $283.0$ & 1 \\
\noalign{\smallskip}\hline\noalign{\smallskip}
\end{tabular}
\end{table*}

Other ocean tide models, such as\footnote{See, e.g., \url{http://amcg.ese.ic.ac.uk/index.php?title=Local:Global_Tidal_Models} on the WEB.}, e.g., GOT99 \cite{got99}, FES2004 \cite{2006OcDyn..56..394L}, CSR $4.0$ \cite{csr40} and EOT11a \cite{2012JGeo...59...28M}, have been produced since EGM96; a comparison between them can yield a measure of the actual uncertainties in $C_{\ell,m,f}^{+},\varepsilon_{\ell,m,f}^{+}$. For example, let us consider the EOT11a model\footnote{Its coefficients in the $\{C_{\ell,m,f}^{+},\varepsilon_{\ell,m,f}^{+}\}$ formalism can be downloaded at \url{http://portal.tugraz.at/portal/page/portal/Files/i5210/files/projekte/COTAGA/oceanTide_eot11a_v20110810_IERS.txt}.} \cite{2012JGeo...59...28M}, used also in \cite{2013arXiv1311.3917M}.
\begin{table*}[ht!]
\caption{Ocean tidal height coefficients $C_{\ell,m,f}^{+}$ for the $f=K_1$ and $f=K_2$ tidal constituents ($\ell=2,p=1,q=0$) from the EGM96 \cite{EGM96} and EOT11a \cite{2012JGeo...59...28M} models and their differences $\left|\Delta C_{\ell,m,f}^{+}\right|$. It must be noted that $\left|\Delta C_{\ell,m,f}^{+}\right|$ are larger than the errors released in the EGM96 model \cite{EGM96}, listed in Table \ref{tavola3}.
}\label{tavola6}
\centering
\bigskip
\begin{tabular}{llll}
\hline\noalign{\smallskip}
 Tide $f$ & $C_{\ell,m,f}^{+}$ (cm) \cite{EGM96} & $C_{\ell,m,f}^{+}$ (cm)  \cite{2012JGeo...59...28M} & $\left|\Delta C_{\ell,m,f}^{+}\right|$ (cm)  \\
\noalign{\smallskip}\hline\noalign{\smallskip}
$K_1$ & $2.83$ & $2.2668$ & $0.56$  \\
$K_2$ & $0.27$ & $0.5355$ & $0.26$ \\
\noalign{\smallskip}\hline\noalign{\smallskip}
\end{tabular}
\end{table*}
\begin{table*}[ht!]
\caption{Ocean phase lag angles $\varepsilon_{\ell,m,f}^{+}$ for the $f=K_1$  and $f=K_2$ tidal constituents ($\ell=2,p=1,q=0$) from the EGM96 \cite{EGM96} and EOT11a \cite{2012JGeo...59...28M} models and their differences $\left|\Delta \varepsilon_{\ell,m,f}^{+}\right|$. It must be noted that $\left|\Delta \varepsilon_{\ell,m,f}^{+}\right|$ are larger than the errors released in the EGM96 model \cite{EGM96}, listed in Table \ref{tavola3}.
}\label{tavola7}
\centering
\bigskip
\begin{tabular}{llll}
\hline\noalign{\smallskip}
 Tide $f$ & $\varepsilon_{\ell,m,f}^{+}$ (deg) \cite{EGM96} & $\varepsilon_{\ell,m,f}^{+}$ (deg)  \cite{2012JGeo...59...28M} & $\left|\Delta\varepsilon_{\ell,m,f}^{+}\right|$ (deg)  \\
\noalign{\smallskip}\hline\noalign{\smallskip}
$K_1$ & $320.6$ & $317.536$ & $3.1$ \\
$K_2$ & $328.4$ & $319.523$ & $8.9$ \\
\noalign{\smallskip}\hline\noalign{\smallskip}
\end{tabular}
\end{table*}
It is remarkable how the differences $\abs{\Delta C^{+}_{\ell,m,f}},\abs{\Delta \varepsilon ^{+}_{\ell,m,f}}$ between the values of EGM96 \cite{EGM96} and EOT11a \cite{2012JGeo...59...28M} are larger than the errors released in EGM96 itself \cite{EGM96}, listed in Table \ref{tavola3}. Thus, if $\abs{\Delta C^{+}_{\ell,m,f}},\abs{\Delta \varepsilon ^{+}_{\ell,m,f}}$ were to be used as realistic uncertainties for $C^{+}_{\ell,m,f},\varepsilon ^{+}_{\ell,m,f}$, the
results of Figure \ref{oceantidalplots} and Table \ref{tavola5} turn out to be optimistic, as confirmed by Table \ref{tavola8}.
\begin{table*}[ht!]
\caption{Minimum and maximum percent ocean tidal bias on the Lense-Thirring node precession for a Galileo satellite. Also the corresponding values for the time span $T$ and the initial value $\Om_0$ of the satellite's node are listed. For  $C_{\ell,m,f}^{+}$ and $\varepsilon_{\ell,m,f}^{+}$,  we adopted the averages of their values by EGM96 \cite{EGM96} and EOT11a \cite{2012JGeo...59...28M}. For their uncertainties, we assumed the differences $\abs{\Delta C^{+}_{\ell,m,f}},\abs{\Delta \varepsilon ^{+}_{\ell,m,f}}$ listed in Table \ref{tavola6} and Table \ref{tavola7}.
}\label{tavola8}
\centering
\bigskip
\begin{tabular}{llll}
\hline\noalign{\smallskip}
Tide $f$ & Min/max percent bias $(\%)$ & $\Om_0$ (deg) & $T$ (yr)  \\
\noalign{\smallskip}\hline\noalign{\smallskip}
$K_1$ (min) & $21$ & $276.3$ & 10 \\
$K_1$ (max) & $986$ & $337.6$ & 1 \\
$K_2$ (min) & $8$ & $29.2$ & 10 \\
$K_2$ (max) & $564$ & $308.5$ & 1 \\
\noalign{\smallskip}\hline\noalign{\smallskip}
\end{tabular}
\end{table*}
\subsection{The inclusion of  Galileo in linear combinations with the LAGEOS satellites}\lb{combiz}
In \cite{2013arXiv1311.3917M}, it is proposed to combine the nodes of the 27 spacecraft of the Galileo constellation with those of the LAGEOS family to reduce the systematic bias due to the mismodeling in the even zonals $J_{\ell}$ of the Earth's geopotential with respect to the past and future tests. Indeed, the authors of \cite{2013arXiv1311.3917M} write:  \virg{The LTE has already been measured by using [\ldots]
the LAGEOS satellites [\ldots]
 with an accuracy of about $10\%$ and will be improved down to a few percent with the recent LARES experiment. The Galileo system will provide 27 new node observables [\ldots]
 and their combination with the LAGEOS and LARES satellites can potentially reduce even more the error due to the mismodeling in Earth's gravity field.}.  In several works it has been pointed out how the total accuracy in the LAGEOS/LAGEOS II performed test is likely larger than $10\%$ by a factor $\sim 2-3$ or so; see \cite{2011Ap&SS.331..351I, 2013CEJPh..11..531R} and references therein. The idea of combining the nodes of the LAGEOS-type satellites with those of existing higher-altitude SLR targets, such as Etalon 1 and Etalon 2 and the GPS spacecraft, was put forth for the first time in \cite{2002JGSJ...48...13I, 2002CQGra..19.5473I, 2005AdSpR..36..472V}.
As a general remark,  the  proposal by the authors of \cite{2013arXiv1311.3917M} seems rather problematic with respect to their declared main goal, and also as far as the overall uncertainty is concerned. Indeed, in the considered scenario, the linear combination approach, reviewed below,  would allow to cancel just the first 29 even zonals at most. From the perspective of some of the authors of \cite{2013arXiv1311.3917M}, it sounds contradictory since they repeatedly claimed that the even zonals of degree higher than $\ell = 6$ would have no impact on a $\sim 1\%$ test including only the three satellites of the LAGEOS family. Thus, it is difficult to understand the need of adding an important source of additional disturbances such as the Galileo spacecraft for nothing. On the other hand, such an optimistic view about the effect of the even zonals on the satellites of the LAGEOS family has been criticized so far with a number of independent and quantitative arguments \cite{2011Ap&SS.331..351I, 2012CaJPh..90..883R} by pointing out that the low altitude of LARES may unfavorably impact an accurate  measurement of the Lense-Thirring effect because of the even zonals of degree as high as $\ell \gtrsim 50-60$. Different a-priori evaluations of the resulting systematic bias performed with different methodologies and levels of approximation do not converge to the desired $1\%$, yielding oscillating figures which, however,  are larger than $\sim 1\%$ by about one order of magnitude or so (see the review \cite{2011Ap&SS.331..351I} and references therein, and the recent analysis in \cite{2012CaJPh..90..883R}). Thus, the choice of introducing the Galileo spacecraft seems questionable also from such a point of view since it would not remove the potentially detrimental even zonals at the price of introducing further sources of systematic errors.

In the linear combination approach\footnote{For its actual origin and use, and to correct the erroneous claim in \cite{2013arXiv1311.3917M} about the $J_2-$free combination used in the tests with the nodes of LAGEOS and LAGEOS II, see \cite{2011Ap&SS.331..351I, 2013CEJPh..11..531R} and references therein.} applied to the nodes, one writes down $N$ equations
\eqi \delta\dot\Om^{(i)} = \mu_{\rm LT}\dot\Om_{\rm LT}^{(i)}+ \sum_{k=1}^{N-1}\ton{\derp{\dot\Om^{(i)}_{J_{2k}}}{J_{2k}}}\delta J_{2k},\ i=1,2,\ldots N  \eqf
for the residual node rates $\delta\dot\Om^{(i)}$  of $N$ satellites as sums of the Lense-Thirring rate, assumed as unmodeled, plus the secular node precessions due to the first $N-1$ even zonals, assumed as mismodeled. Then, the resulting algebraic linear system of $N$ equations in the $N$ unknowns
\eqi
\underbrace{\mu_{\rm LT}, \delta J_2, \delta J_4 \ldots \delta J_{2(N-1)}}_{N}
\eqf is solved for the parameter $\mu_{\rm LT}$, which is equal to one in GTR. Further algebraic manipulations allow one to obtain the following linear combination
\eqi {\mathcal{C}}_N \doteq \delta\dot\Om^{(1)} + \sum_{j=1}^{N-1}c_j\delta\dot\Om^{(j+1)} \lb{combo}\eqf
of the $N$ residual node rates which, by construction, is independent of the first $N-1$ even zonals, being impacted by the other ones of degree $\ell > 2(N-1)$ along with the non-gravitational perturbations and other possible orbital perturbations which cannot be reduced to the same formal expressions of the first $N-1$ even zonal rates. It turns out that $\mu_{\rm LT}$ can be expressed as the ratio of \rfr{combo} to
\eqi\mathcal{C}_{\rm LT} \doteq \dot\Om_{\rm LT}^{(1)} + \sum_{j=1}^{N-1}c_j\dot\Om_{\rm LT}^{(j+1)} \lb{comboLT}.\eqf
The coefficients $c_j,\ j=1,2,\ldots N-1$ in \rfr{combo}-\rfr{comboLT} depend only on the semimajor axes $a^{(i)}$, the eccentricities $e^{(i)}$, and the inclinations $I^{(i)}$ of the $N$ satellites involved in such a way that, by construction, ${\mathcal{C}}_N=0$ if \rfr{combo} is calculated by posing \eqi\delta\dot\Om^{(i)}=\ton{\derp{\dot\Om^{(i)}_{J_{\ell}}}{J_{\ell}}}\delta J_{\ell},\ i=1,2,\ldots N\eqf for any of the first $N-1$ even zonals, independently of the value assumed for its uncertainty $\delta J_{\ell}$. It should be stressed that \rfr{combo} vanishes if calculated for any effect having the same functional form of  $\partial\dot\Om^{(i)}_{J_\ell}/\partial J_{\ell},\ i=1,2,\ldots N$ for given values $\ell=2,4\,\ldots 2(N-1)$; it is the case of the node variations due to secular, seasonal and harmonic  time-dependent components of $J_{\ell}$ for  any of $\ell=2,4\,\ldots 2(N-1)$.

The previous discussion shows that the proposal of using the nodes of more than one Galileo satellites is, in principle, unfeasible. Indeed, all the Galileo spacecraft will share nominally  the same semimajor axis $a$, eccentricity $e$ and inclination $I$, entering the node classical precessions $ \dot\Om_{J_{\ell}}$ which the coefficients $c_j,\ j=1,2,\ldots N-1$ are made of.
Thus, in principle, only one Galileo spacecraft could be included in linear combinations with the satellites of the LAGEOS family.

Also invoking the fact that, in practice, small differences in the orbital parameters from satellite to satellite will occur because of unavoidable orbit injection errors; these would not substantially alter the unfavourable picture. Indeed, if, on the one hand, the linear combination approach would formally work fine in canceling the impact of the first $N-1$ even zonals, on the other hand, the coefficients of the Galileo nodes would be quite large, thus enormously enhancing the overall impact of the uncancelled Galileo non-gravitational and tidal $K_1$, $K_2$ perturbations on the combination.
For example, let us include a pair of navigation spacecraft, dubbed G1, G2, in a $N=5$ combination including also LAGEOS (L), LAGEOS II (L II), LARES (LR); let it be
\eqi \mathcal{C}_5 \doteq \delta\dot\Om^{(\rm L)} + c_1\delta\dot\Om^{(\rm L\ II)} + c_2\delta\dot\Om^{(\rm LR)} + c_3\delta\dot\Om^{(\rm G1)} + c_4\delta\dot\Om^{(\rm G2)}. \eqf
By assuming a difference of, say, $\Delta a=10$ km in the semimajor axes of G1 and G2,
their coefficients would amount to
\begin{align}
c_3 & = -40.3, \\ \nonumber \\
c_4 & = 110.8.
\end{align}

Let us, now, look at only one Galileo spacecraft.
%
In this case, it would be possible to cancel out the bias due to the mismodeling in the first three even zonals $J_2,J_4,J_6$.
Table \ref{tavola9} lists the partial derivatives of their classical node precessions with respect to the even zonals themselves for the satellites involved.
They can be calculated by using the analytical results in \cite{2003CeMDA..86..277I}, valid up to $\ell=20$.
\begin{table*}[ht!]
\caption{Coefficients $\partial{\dot\Om_{J_{\ell}}}/\partial J_{\ell}$, in mas yr$^{-1}$, of the satellites of the LAGEOS family and of a typical Galieo-type spacecraft.
In order to have either the nominal or the mismodelled node precessions of degree $\ell$, they must be multiplied either by $J_{\ell}$ or by $\delta J_{\ell}$, respectively.
}\label{tavola9}
\centering
\bigskip
\begin{tabular}{lllll}
\hline\noalign{\smallskip}
$\ell$ & LAGEOS & LAGEOS II & LARES & Galileo \\
\noalign{\smallskip}\hline\noalign{\smallskip}
2 & $4.17159\times 10^{11}$ & $-7.66948\times 10^{11}$ & $-2.06930\times 10^{12}$ & $ -3.14280\times 10^{10} $  \\
4 & $1.54225\times 10^{11}$ & $-5.58677\times 10^{10}$ & $-1.83868\times 10^{12}$ & $ -7.39756\times 10^{8} $  \\
6 & $3.27732\times 10^{10}$ & $4.99242\times 10^{10}$ & $-9.06244 \times 10^{11}$ & $ 4.27652\times 10^{7} $  \\
\noalign{\smallskip}\hline\noalign{\smallskip}
\end{tabular}
\end{table*}
In Table \ref{tavola10} we display the coefficients of the nodes of LAGEOS, LAGEOS II, LARES, Galileo for the 24 different linear combinations ${\mathcal{C}}_4$ which can be constructed with them along with the corresponding combined Lense-Thirring trends.
\begin{table*}[ht!]
\caption{Weighting coefficients  of the nodes of LAGEOS, LAGEOS II, LARES, Galileo entering the 24 different linear combinations ${\mathcal{C}}_4$ which can be constructed with them, and the corresponding combined Lense-Thirring trends ${\mathcal{C}}_{\rm LT}$. The coefficients displayed here are made of $\partial\dot\Om_{\ell}/\partial J_{\ell}, \ell=2,4,6$ listed in Table \ref{tavola9} according to the linear combination approach described in the text.
}\label{tavola10}
\centering
\bigskip
\begin{tabular}{llllll}
\hline\noalign{\smallskip}
& LAGEOS & LAGEOS II & LARES & Galileo & ${\mathcal{C}}_{\rm LT}$ (mas yr$^{-1}$)\\
\noalign{\smallskip}\hline\noalign{\smallskip}
1) &  $  1  $  &  $  0.587464  $  &  $ 0.0682644  $  &  $  -5.5573  $  &  $ 45.0817 $  \\
2) &  $  1  $  &  $  0.587464  $  &  $  0.501944  $  &  $  -0.755793  $   &  $  106.774  $ \\
3) &  $  1  $  &  $  0.303317  $  &  $  0.132214  $  &  $  -5.5573  $  &  $  43.686  $ \\
4) & $  1  $  &  $  0.303317  $  &  $  0.501944  $  &  $  -1.46382  $  &  $  96.2815  $ \\
5) & $  1  $  &  $  2.23027  $  &  $  0.132214  $  &  $  -0.755793  $  &  $  114.844  $  \\
6) & $  1  $  &  $  2.23027  $  &  $  0.0682644  $  &  $  -1.46382  $  &  $ 105.747 $ \\
7) & $  1.65809  $  &  $  1  $  &  $  0.114685  $  &  $  -9.33634  $  &  $  75.4764  $  \\
8) & $  1.65809  $  &  $  1  $  &  $  0.843273  $  &  $  -1.26974  $  &  $ 179.121 $  \\
9) & $ 0.442523  $  &  $  1  $  &  $  0.0600657  $  &  $  -0.338879  $  &  $ 51.4063  $  \\
10) & $  0.442523  $  &  $  1  $  &  $  0.0306081  $  &  $  -0.665021  $  &  $  47.2158  $ \\
11) & $  3.25385  $  &  $  1  $  &  $  0.44166  $  &  $  -18.3218  $  &  $  143.389  $   \\
12) & $  3.25385  $  &  $  1  $  &  $  1.65485 $  &  $ -4.88986  $  &  $  315.97  $ \\
13) & $ 3.85417  $  &  $  4.50203 $  &  $  1  $  &  $ -42.0325  $  &  $  286.203  $ \\
14) & $ 1.01521  $  &  $  1.18586  $  &  $  1  $  &  $  -2.9163  $  &  $  180.171  $ \\
15) & $  1.01521  $  &  $  1.17038  $  &  $  1  $  &  $  -2.95487  $  &  $  179.599  $ \\
16) & $  3.80386 $  &  $  4.38528 $  &  $  1  $  &  $ -41.4839  $  &  $  282.183  $ \\
17) & $  3.80386 $  &  $ 16.6484  $  &  $  1  $  &  $ -10.927  $  &  $  735.032  $ \\
18) & $  3.85417  $  &  $  16.8686  $  &  $  1  $  &  $  -11.218  $  &  $  742.871  $ \\
19) & $  -4.95756 $  &  $  -0.787562  $  &  $  -0.174935  $  &  $  1  $  &  $  -195.273  $ \\
20) & $  -2.55967 $  &  $  -0.40663  $  &  $  -0.0466344 $  &  $  1  $  &  $  -94.608  $ \\
21) & $  -2.55967 $  &  $  -0.207209 $  &  $  -0.091516 $  &  $  1  $  &  $  -93.6284 $ \\
22) & $ -2.52626  $  &  $  -0.204505 $  &  $ -0.0891427  $  &  $  1  $  &  $  -92.2387  $ \\
23) & $  -4.95756  $  &  $  -0.777282 $  &  $  -0.177248 $  &  $  1  $  &  $  -195.223  $ \\
24) & $  -2.52626 $  &  $  -0.396085 $  &  $  -0.0460257 $  &  $  1  $  &  $  -93.1797  $  \\
\noalign{\smallskip}\hline\noalign{\smallskip}
\end{tabular}
\end{table*}
While all of them are, by construction, equivalent from the point of view of the systematic uncertainty due to the uncanceled even zonals $J_8, J_{10},\ldots$, it is not so, in principle, as far as the impact of the other uncancelled orbital perturbations of gravitational and non-gravitational origin is concerned. To this aim, dedicated analyses are required. Let us make just a simple order-of-magnitude evaluation. At first glance, the combination 9) of Table \ref{tavola10} seems to be the most promising because of the smallness of its coefficients. According to Figure 5 and Figure 7 of \cite{2013arXiv1311.3917M}, the Solar radiation pressure (SRP) may impact the node of a Galileo satellite with rates up to $\approx 4-60$ times larger than the Lense-Thirring one, depending on the model parameters which are to be estimated. Let us assume a SRP residual node rate for Galileo as large as, say,
\eqi\dot\Om^{(\rm G)}_{\rm SRP}=10\dot\Om_{\rm LT}^{(\rm G)}=22\ {\rm mas\ yr^{-1}}.\lb{srp}\eqf
By inserting \rfr{srp} in the combination 9) of Table \ref{tavola10}, a total bias of $11\%$ occurs. If the combination 1) of Table \ref{tavola10} is considered, the resulting percent error is $270\%$. It reaches $323\%$ with the combination 13) of Table \ref{tavola10}.
\section{Conclusions}\lb{fine}
The recent proposal by Moreno Monge et al. of combining the data of all the members of the planned Galileo constellation of navigation spacecraft with those of the existing satellites of the LAGEOS family to increase the overall accuracy of the performed and ongoing Lense-Thirring tests has several drawbacks.

A first issue is represented by the actual measurability of the sought gravitomagnetic effect in the Galileo station-satellite SLR range. Indeed, its peak-to-peak amplitude is as small as $0.5$ mm over 3-days orbital arcs. Such a figure has to be compared to the $\approx 1-15$ cm-level RMS accuracy reached in the orbit determination of the tests satellites GIOVE-A and GIOVE-B.  It would take 15 years for the gravitomagnetic field to reach a m-level  peak-to-peak  range amplitude with one Galileo spacecraft.

Major issues come also from the systematic errors of some gravitational and non-gravitational orbital perturbations.

Using only one Galileo spacecraft would be impossible because of the aliasing due to the first even zonal harmonic of the multipolar expansion of the geopotential. Indeed, the mismodeling in its static component would cause a residual node precession as large as $30-300\%$ of the gravitomagnetic one.
Another major issue is represented by those long-period harmonic tidal orbital perturbations, such as  those due to the $K_1$ and $K_2$ constituents, having frequencies which are multiple of the Galileo's node rate itself. Indeed, the orbital plane of a Galileo satellite will take about 40 years to complete a full cycle. Thus, the $K_1$ and $K_2$ node perturbations would act as superimposed semisecular aliasing trends over necessarily limited observational time spans $T$. It turns out that their mismodeling may impact the Lense-Thirring  node precession of a Galileo satellite up to $300\%$, depending on $T$ and on the initial value $\Om_0$ of the node.

The linear combination approach, designed to remove the impact of the first $N-1$ even zonals by suitably combining the nodes of $N$ satellites with different orbital configurations, does not allow, in principle, to include  the entire Galileo constellation. Indeed, all its members will have nominally the same semimajor axis $a$, eccentricity $e$ and inclination $I$; the multiplicative coefficients of the nodes entering the linear combinations depend only on $a,e,I$. On the other hand, using their data separately would fully expose them to the impact of all the perturbations previously mentioned.
It could be replied that, actually, the unavoidable orbital injection errors will prevent to have exactly the same orbital configurations for all the Galileo spacecraft. Nonetheless, even small differences in their orbital parameters would yield large values of the coefficients weighting their nodes in the global linear combination, thus enhancing the biasing impact of the uncancelled perturbations such as the tesseral and sectorial tides and the non-gravitational effects. In principle, one could adopt one of the 24 independent combinations involving the nodes of only one Galileo satellite and those of the existing three members of the LAGEOS family. Actually, it would not be useful since the coefficient of the Galileo's node would be, in any case, too large. In the potentially most favorable combination, the impact of the Galileo's node rate induced by the Solar radiation pressure would be as large as $11\%$ of the Lense-Thirring combined trend.

All in all, it is difficult to understand why  a further, serious source of non-negligible systematic uncertainties like one or more Galileo satellites should be added for essentially nothing. This is true either if one trusts the claims by Ciufolini et al. about the irrelevance of all the even zonals of degree higher than 6 in the ongoing LAGEOS-LAGEOS II-LARES test and if, instead, one gives credit to people warning  about the potentially detrimental effect of the even zonals of degree $\gtrsim 50-70$ due to the relatively low altitude of LARES. Indeed, even if one could finally combine all the Galileo spacecraft with the LAGEOS-type satellites, the impact of only the first 29 even zonals, most of which do not represent an issue for a LAGEOS family test according to both the opposite positions previously outlined, would be removed.

\bibliography{Galileobib}{}

\end{document}